\documentclass[a4paper,fleqn,usenatbib,useAMS]{mnras}
\usepackage{newtxtext,newtxmath}
\usepackage[T1]{fontenc}
\usepackage{ae,aecompl}
\usepackage{graphicx}    
\usepackage{amsmath}     
\usepackage{amssymb}     
\usepackage{multicol}    
\usepackage{bm}          
\usepackage{pdflscape}   
\usepackage{xspace}
\newcommand{\msun}{\,$M_{\sun}$}

\newcommand{\rsun}{\,$R_{\sun}$}
\newcommand{\ergs}{\,erg\,s$^{-1}$}
\newcommand{\kms}{\,km\,s$^{-1}$}

\newcommand{\heii}{\,He\,II}
\newcommand{\ha}{\,H$\alpha$}
\title[The explosion energy of SN~2013fs] 
  {The explosion energy of the type IIP supernova SN~2013fs with a confined dense circumstellar shell}
\author[N. N. Chugai]{
N. N. Chugai\thanks{E-mail: nchugai@inasan.ru}
\\
$^{1}$Institute of Astronomy, Russian Academy of Sciences, Pyatnitskaya
      St. 48, 119017 Moscow, Russia\\
}

\date{Accepted XXX. Received YYY; in original form ZZZ}

\pubyear{2019}

\begin{document}
\label{firstpage}
\pagerange{\pageref{firstpage}--\pageref{lastpage}}
\maketitle
%
\begin{abstract}

The recent study of SN~2013fs flash spectrum suggests enormous for SN~IIP 
    explosion energy, far beyond possibilities of the neutrino mechanism. 
The issue of the explosion energy of SN~2013fs is revisited making use of 
    effects of the early supernova interaction with the dense circumstellar shell.
The velocity of the cold dense shell between reverse and forward shocks is 
    inferred from the analysis of the broad \heii\,4686\,\AA\ on day 2.4.
This velocity alongside with other observables provide us with an alternative 
   energy estimate of $\sim1.8\times10^{51}$\,erg for the preferred mass of $\sim10$\msun.
The inferred value is within the range of the neutrino driven explosion.
 
\end{abstract}

\section{Introduction} 
\label{sec:intro}
A single star with an initial mass of 9-25\msun\ retains significant amount of 
 the hydrogen envelope prior to the core collapse and explodes as a red supergiant 
 (RSG) giving rise to SN~IIP \citep{HFWLH_03} ("P" stands for the light curve "plateau") or 
 SN~IIL ("L" for linear light curve). 
In some cases a distinction between  SN~IIP and SN~IIL is illusive, so SN~II as 
  a designation SN~II becomes more adequite.
A RSG pre-SN loses its mass via the slow wind with velocities reliably known 
  only for a few objects, e.g., $u_w = 30$\kms\ for SN~1998S \citep{Chu_02} 
  and $u_w = 20$\kms\ for SN~1997eg \citep{Chu_19}. 
The typical wind density parameter is $w = \dot{M}/u \sim 10^{14}-10^{15}$\,g\,cm$^{-1}$ as 
  inferred from radio and X-ray data \citep{CFN_06}. 
The SN interaction with the wind 
  of that density cannot produce detectable optical emission lines. 
  
Yet following SN~2006bp \citep{Quimby_07} at least dozen of SNe~II show 
 high ioniation emission lines on a smooth continuum in early ($< 10$ days) 
  spectra \citep{Khazov_16}.
This phenomenon dubbed "flash spectrum" is the emission lines from a confined 
  dense circumstellar (CS) shell ($r \lesssim 10^{15}$ cm) ionized by 
  the flash of UV radiation after the shock breakout 
  \citep{Groh_14,Khazov_16,Yaron_17}. 
 The phenomenon of high ionization emission lines in early spectra was observed 
   formerly in the type IIL SN~1998S \citep{Fassia_01} 
   and interpretted as originated in the confined ($r \lesssim 10^{15}$ cm) CS shell 
   with the mass of $\sim 0.1$\msun\ and the Thomson optical depth of $\sim 2-4$ \citep{Chu_01}.
The mechanism of the vigorous mass loss responsible for 
  the dense confined CS shell is not yet understood.
Only $\sim 15$\% of SNe~II demonstrate signatures of the confined CS shell   
  \citep{Khazov_16} which suggests that some special preconditions 
  (e.g., initial mass, rotation)
  should be met for the  masive star to become "flash" SN~II.

Recently, \citet{Yaron_17} explored flash spectrum of SN~2013fs 
   and among important results on the structure of the confined CS shell authors 
  reported an estimate of the explosion energy based on the analysis 
  of the early (< 3 day) multiband photometry. (Here we identify the explosion energy and the ejecta kinetic energy.)  
For the ejecta mass of 10\msun\ the inferred energy is amazingly high
  $\approx 5\times10^{51}$ erg   \citep{Yaron_17}.
Their estimate significantly exceeds the explosion energy for normal SNe~IIP and  
  is also far above the upper limit  ($2\times10^{51}$ erg)   
  for the neutrino explosion mechanism \citep{Janka_17}.
Although supernovae could be caused by the more energetic explosions (e.g.,   
  hypernovae), it seems that currently it is premature to abandon the neutrino mechanism for SN~2013fs.
A hint for a possible uncertainty of the energy estimate stems from a significant  
  discordance of the RSG radius inferred by the hydrodynamic and analytical approaches 
  \citep{Yaron_17}.
In this situation an alternative estimate of the explosion energy of SN~2013fs is 
  highly demanded.

Here we explore the early stage of the hydrodynamic interaction of SN~2013fs ejecta 
  with the dense CS shell in  an attempt to propose an alternative energy estimate.
Particularly we make use of the broad emission at about 4500\,\AA\ at 2.4 day 
   identified with the \heii\,4686\,\AA\ line \citep{Bull_18}.
It will be argued that this line originates from the fragmented cold dense shell (CDS) between 
  the reverse and forward shock (Section \ref{sec:he}).
The inferred CDS velocity in a combination with other observables will permit us to constrain the 
  explosion energy via the modelling of the SN/CSM interaction (Section \ref{sec:energy}).

The study is based on the Keck-1 spectrum at the age of 2.42 days
  (Oct. 8.45) and the spectrum on day 51 (Nov. 26) taken at the 
  Palomar 200-inch telescope; both spectra are 
   retrieved from the WISeREP database \citep{Yaron_12} ({\it https://wiserep.weizmann.ac.il}). 
Below I use the explosion date Oct. 6.03 (UTC), the average between 
  2013 Oct. 6.12 \citep{Yaron_17} and Oct. 5.946 \citep {Bull_18}.

 %
 \begin{figure}
 	\includegraphics[trim= 30 60 0 0,width=0.9\columnwidth]{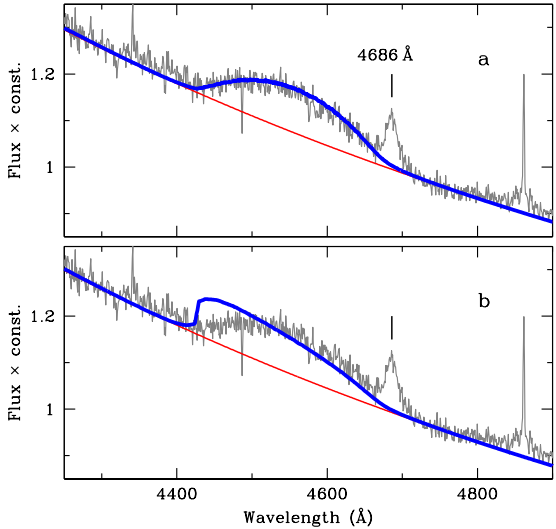}
 	\caption{%
 		The calculated \heii\,4686\,\AA\ line ({\it blue line}) 
 		overplotted on the observed spectrum at 2.42 d ({\it grey}). 
 		The panel {\it a} shows the case of the well mixed CDS ($A \gg 1$), 
 		while the panel {\it b} corresponds to the moderate mixing 
 		($A \sim 1$). The {\it red line} is the adopted continuum.
    	}
 	\label{fig:he}
 \end{figure}
 %
 
 \section{Broad \heii\,4686\,\AA} 
\label{sec:he}

The SN~2013fs spectrum on day 2.42 shows a broad low contrast 
  emission at about 4500\,\AA\ \citep{Bull_18}. 
Authors identify it with \heii\,4686\,\AA\ emitted by the SN ejecta with 
  the velocity of $\gtrsim 15000$\kms.
While the identification seems reasonable, the suggested line-emitting site 
  raises a problem.
The point is that the line-emitting layer in the early unshocked ejecta 
  ionized by $\sim 1$ keV X-rays from the reverse shock is very narrow, 
  $\Delta r/r \sim (k_x\rho r)^{-1} \sim 10^{-2}-10^{-1}$ (where 
  $k_x = 100E_{keV}^{-8/3}$\,cm$^2$\,g$^{-1}$ is 
  the X-ray absorbtion coefficient) and accordingly
  has a small velocity dispersion $\Delta v/v = \Delta r/r \ll 1$. 
The line profile in this case should be boxy, skewed towards blue by 
  the occultation effect. 
The observed broad \heii\ line is indeed skewed towards blue, but the original 
   profile looks dome-like, not boxy \citep[cf.][Fig. 20]{Bull_18}.

An alternative line-emitting site is the CDS, although in this case the 
  dome-like profile requires special preconditions. 
The spherical thin shell predicts M-shaped profile for the optically 
  thick line \citep{Gerasim_33,Cid_94} and boxy profile for the optically 
  thin line, both options inconsistent with the observed profile.  
In fact, a smooth thin CDS cannot survive. The decelerating 
  CDS is subject to the Rayleigh-Taylor instability 
  that results in the CDS fragmentation, which produces  
  an ensemble of filaments and randomly folded sheets of the dense 
  cold gas \citep{CheBlo_95,BloEll_01}. 
The line profile emitted by the fragmented shell for the optically thin line  
  is boxy.
Moreover, for the optically thick line the profile may also be boxy. 
The profile shape depends on the mixing degree specified by the area ratio $A = S/4\pi R^2$, i.e.,
  the ratio of the cumulative surface $S$ of CDS fragments to the undisturbed 
  CDS surface. 
The parabolic profile is expected for $A \gg 1$ and boxy profile for 
  $A \sim 1$ \citep{CCL_04}. 
The dome-like profile of the \heii\ line at 2.4 days suggests that the CDS should be fragmented 
  and well mixed. 
A possible clumpy structure of the CS shell additionally favours the advanced mixing 
  of the CDS \citep[cf.][Fig. 6]{Blondin_01}.

The computed \heii\,4686\,\AA\ profile is shown for two extreme mixing cases, 
  $A \gg 1$ and $A\sim 1$ (Fig. \ref{fig:he}). 
The fragmented CDS in the model lies at the sharp photosphere that presumably 
   coincides with the inner boundary of the CDS.
 The fragmented shell should absorb in the continuum in order to fit the profile 
  around zero and positive radial velocities.
The optimal optical depth is 0.2. 
The adopted Thomson optical depth of the wind is $\tau_{\textsc{T}} = 1$ in line with 
   the  estimates of \citet{Yaron_17}.
The result is not sensitive to the variation of the wind Thomson optical depth 
  in the range of $\pm 30$\%.
The case $A \gg 1$ provides an excellent fit while the  case $A \sim 1$
  is rulled out. 
The expansion velocity of the CDS is found to be $16600\pm300$\kms\ that is 
  consistent with the \citet{Bull_18} estimate of the velocity of the line-emitting 
  gas ($\gtrsim 15000$\kms).
The narrow \heii\,4686\,\AA\ line with the FWHM $\sim 10^3$\kms\ (Fig. \ref{fig:he}) 
   that is apparently emitted by the CS gas will be  discussed in Section \ref{sec:disc}.

The broad \heii\,4686\,\AA\ emission is likely dominated by the recombination. 
In that case for the solar composition of the hydrogen and helium  one expects 
  the comparable emission of the \ha\ that is actually not seen on day 2.4  
  \citep[cf.][Fig. 7]{Bull_18}. 
The likely reason for that is a significant He enrichment indicated by the low    
  ratio  \ha/\heii\,4686\,\AA\ $\sim 0.3$ in the CS flash spectrum of SN 2013fs \citep{Yaron_17}.
   
%
\begin{figure}
   \includegraphics[trim=50 40 0 0,width=0.95\columnwidth]{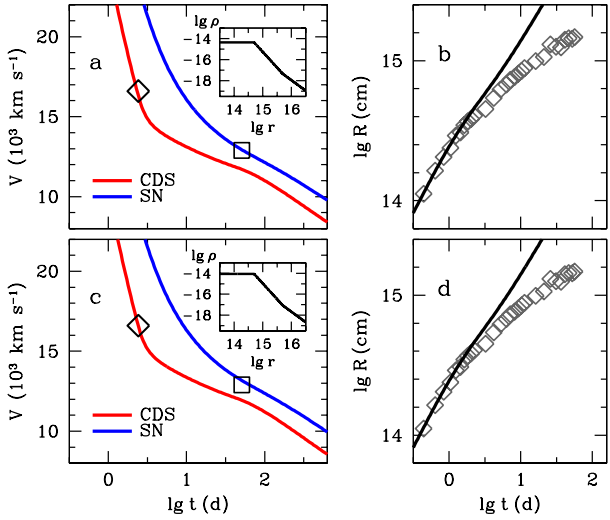}
   \caption{%
   The SN/CSM interaction model compared to observations for the low 
     energy model A (panel {\it a, b}) and high energy mosel B (panels {\it c, d}).
   The panel {\it a} shows  the model CDS velocity ({\it red}) vs. time  
      compared to the inferred CDS velocity at 0.52 d ({\it diamond}) 
      and the terminal velocity of the unshocked SN ejecta ({\it blue}) 
      compared to the terminal velocity 
      of the unshocked SN ejecta on day 51 recovered from the \ha\
       ({\it square}); the inset shows the CS density 
        distribution (CGS units). 
   The panel {\it b} shows the CDS radius vs. time 
     compared to the photosphere radius ({\it diamonds}) \citep{Yaron_17}. 
   The panels {\it c} and {\it d} are the same as {\it a} and {\it b} but 
       for the model B.
    }
   \label{fig:maxv}
\end{figure}
%

\section{CS interaction and the explosion energy}
\label{sec:energy}

\subsection{Kinematic constraints}
\label{sec:kinem}

The inferred expansion velocity of the CDS  can be used alongside with 
  the maximal velocity of the unshocked ejecta $v_{max} = 12900\pm300$\kms\  at $t = 51$ d
  recovered from the blue edge of the \ha\ absorption in the Palomar spectrum
  to constrain the  explosion energy of SN~2013fs for a given ejecta mass. 
An important additional constraint is provided by the photospheric radius  
   \citep{Yaron_17} that should coincide with the model CDS at the early stage.  
In order to describe these observables we consider the SN interaction with the 
  CS gas in the thin shell approximation \citep{Che_82,Giu_82,Chu_01}.
Yet, as we will see, kinematic considerations only are not able to constrain    
   the energy. 
 The point is that the change of the energy can be compensated by a corresponding 
   change of the  CS density so that the CDS velocity at the certain age is preserved.
In the next section models with different energy and CS density will be discriminated  
   using the \ha\ liminosity.
  
The dynamic evolution of the CDS is determined by the density distribution of SN   ejecta 
in outer layers $\rho(v)$ and the CS density $\rho_{cs}(r)$.
The density of the homologously expanding ejecta is set as $\rho = \rho_0/[1 + (v/v_0)^q]$ 
  with $\rho_0$ and $v_0$ determined by the ejecta mass $M$ and kinetic energy $E$. 
The hydrodynamic model of SN~2018in (the normal SN~IIP) suggests 
  $q = 7.6$ \citep{UC_13} that is adopted here. 
The density distribution of the confined CS shell $\rho_{cs}(r)$ is presumably 
  uniform within the extent of the confined CS shell $5\times10^{14}$\,cm, 
   the value estimated by \citet{Yaron_17}.  
For  $r > 5\times10^{14}$\,cm the CS density presumably falls as $\rho \propto r^{-3}$ 
  and beyond $5\times10^{15}$\,cm the CS density turns to a slow RSG wind with 
  $\rho \propto r^{-2}$ and the density parameter  $w = \dot{M}/u_w \sim 10^{15}$\,g\,cm$^{-1}$.

For the fiducial model A we adopt the kinetic energy of $1.8\times10^{51}$\,erg,  
 very mush close to the upper limit for the neutrino mechanism. 
In the model B the energy is set to be $3\times10^{51}$\,erg\ that 
   by a factor of 1.5 exceeds the upper limit for the 
    neutrino mechanism.
In both models the ejecta mass is $M = 10$\msun\ in line with  
  the value inferred by \citet{Yaron_17}.
Yet we explore other possibilities as well.
The CS density for the same normalized distribution is determined for each model 
  from the optimal fit (Fig. \ref{fig:maxv}). 
Both presented models reproduce the CDS velocity at $t = 2.42$ d, 
  maximal velocity of unshocked ejecta at $t = 51$ d, 
  and the photosphere radius at  $t \leq 3$ days (Fig. \ref{fig:maxv}) 
  for the mass of the confined CS shell of $3.6\times10^{-3}$\msun\ in the model A and 
  of $7\times10^{-3}$\msun\ in the model B. 
For other choice of the ejecta mass and the same CS density, the appropriate  
  energy obeys the scaling relation $E \propto M^{0.55}$ that is obtained numerically.

\subsection{Constraint from the \ha\ luminosity}

Both the model A and the model B with the energy beyond the neutrino mechanism
  meet observational requirements.
We therefore need to apply an additional tool to discriminate between the models.
The significant difference of the mass of the confined CS shell in these models 
  prompts that the \ha\ emission from the CS gas could provide us
  a crucial test. 
  
The observed \ha\ luminosity at $t = 1.4$\,d is $L_{32} = 1.94\times10^{39}$\ergs\ 
  \citep{Yaron_17}. 
The emission measure at this age is of $3.9\times10^{63}$ cm$^{-3}$ in the
  model A and  of $4\times10^{64}$ cm$^{-3}$ in the model B.
The efficient recombination  coefficient for the \ha\ emission in the recombination 
  case C (optically thick Balmer lines) 
   is $\alpha_{32} = 1.83\times10^{-13}(T_e/10^4\,K)^{-0.73}$\,cm$^3$\,s$^{-1}$  \citep{Oster_89}.
 The effective temperature at this age is $T_{eff} \approx 15000$\,K  
    \citep{Yaron_17}.
 Since the electron temperature in the preshock wind $T_e < T_{eff}$ 
   and decreases with the radius \citep{Dessart_17} $T_e =10^4$\,K 
   is a resonable estimate.
 For this value we obtain then  $L_{32} = 2.1\times10^{39}$\ergs\   
   in the model A and $L_{32} = 2\times10^{40}$\ergs\ in the model B.
 The \ha\ luminosity in the model A coincides with the observational value, 
   whereas for the model B the \ha\ luminosity is by a factor of 10 higher.
The \ha\ test thus strongly disfavours models with the energy significantly 
 exceeding $2\times10^{51}$ erg.  

%
\begin{figure}
   \includegraphics[trim=50 40 0 0,width=0.95\columnwidth]{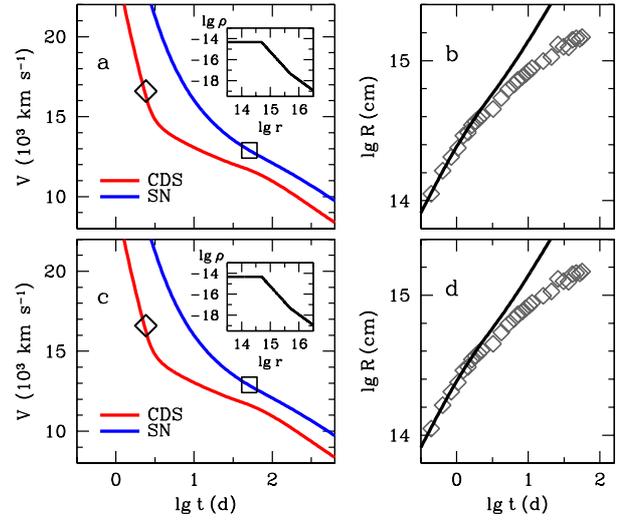}
   \caption{%
  The same as Figure \ref{fig:maxv} but for the ejecta mass of 8\msun\  and 
   the energy of $1.6\times10^{51}$ erg ({\it a, b}) and 
   the ejecta mass of 12\msun\  and the energy of $2\times10^{51}$ erg ({\it c, d}).  
    }
   \label{fig:maxv2}
\end{figure}
%

%
\begin{figure}
	\includegraphics[trim=70 120 -10 0,width=1.0\columnwidth]{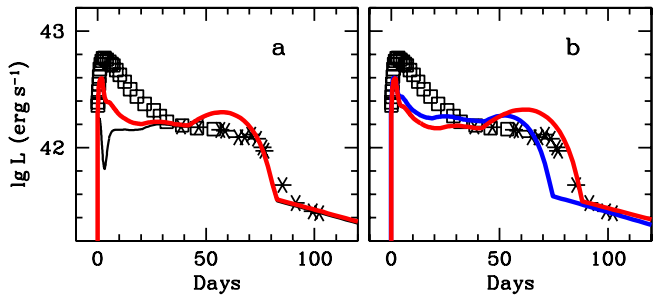}
	\caption{%
   The light curve models overploted on the bolometric light curve ({\em squares}) 
		according \citep{Yaron_17}. {\em Asterisk} symbols is the $R$ band photometry 
		\citep{Bull_18} matched to the bolometric light curve. 
	The panel {\it a} shows the model A without CS interaction ({\em blue}) and 
	   with the CS interaction ({\em red}). 
	The panel {\it b} shows the light curves with the CS interaction for the 
	  low mass ({\em blue}) and high mass ({\em red}) cases.
	}
	\label{fig:lcurve}
\end{figure}
%

\subsection{Ejecta mass and the light curve}

The question may be posed whether the adopted ejecta mass of 10\msun\ is 
  an optimal one.
This estimate is obtained from the anlysis of the shock breakout continuum 
  employing the radiation hydrodynamics \citep{Yaron_17}.  
An independent verification of this choice can be made based on the bolometric light curve.
The observational bolometric light curve reported by \citet{Yaron_17} is truncated 
  at 56 days. 
The sensible reconstruction of the later behavior can be performed with the 
  $R$-band photometry \citep{Bull_18} that we match with the bolometric light curve 
  in the range of 38-56 days (Fig. \ref{fig:lcurve}).
  
The model bolometric light curve is calculated below for the fiducial case A 
  ($M = 10$\msun, $E = 1.8\times10^{51}$ erg), and two alternative cases:
  low mass (8\msun, $E = 1.6\times10^{51}$ erg) and high mass 
  (12\msun, $E = 2\times10^{51}$ erg).  
It should be emphasised that both latter versions meet the same kinematic 
  constraints as the model A (Fig. \ref{fig:maxv2}).   
The ejecta luminosity without the CS interaction is calculated based on a
  simple model that is composed by the two luminosity regimes: 
   the initial diffusion luminosity of the uniform ejecta \citep{Arnett_80}
   and the late time cooling wave  \citep{Chu_91} with the effective temperature of 5000\,K \citep{GIN_71}.
 The cooling wave regime turns on when the diffusion luminosity becomes 
   lower than the cooling wave luminosity; this occurs at about 40-50 days. 
The initial condition is the uniform sphere of the radius $R_0$ 
  with the kinetic energy $E$, the homologous kinematics, and the thermal 
  radiation $E_r = 0.5E$. 
The thermal energy evolution is controlled by the adiabatic cooling,   
  radioactive  heating, and diffusion luminosity $E_r(t)/t_e$. 
The average escape time $t_e$ for the uniform distribution of the matter and 
  radiation density is  $0.2(R/c)\tau$ \citep{ST_80},  where $R$ is the current 
  envelope radius, $c$ is speed of light, and $\tau$ is the optical depth 
  calculated with the Opal opacity \citep{Igles_96}. 

The computed light curve for the model A (Fig. \ref{fig:lcurve}a) is 
  shown for two versions:  without and with the CS interaction.
The interaction luminosity is calculated already in the frame of the CS interaction   
   model (Section \ref{sec:kinem}).
Apart from the adopted mass and energy the ejecta light curve is specified also by the 
  RSG radius of $2.2\times10^{13}$\,cm, the $^{56}$Ni mass of 0.052\msun, and the CS mass  0.0036\msun.
The inferred RSG radius is within the range of previous estimates 
  $(1.2-6.9)\times10^{13}$\,cm  \citep{Yaron_17}   
  
The puzzling minimum in the ejecta luminosity at $\sim 3$ days is related to the maximum 
  in the behavior of the Opal opacity vs. temperature at about $T \approx 1.5\times10^5$\,K.
The minimum becomes pronounced in this particular case due to the approximation of the  
   uniform envelope. 
The significant deficit of the model luminosity at the early epoch is also related to   
  the approximation of the  uniform ejecta: the latter is not able to describe the rapid radiation diffusion expected for real ejecta with rarefied density of outer layers.  
Yet the model satisfactorily reproduces the luminosity and duration of the light curve 
  plateau. 

The light curves for the lower and higher mass (Fig. \ref{fig:lcurve}b) are 
  less successful in the plateau description although the difference with 
  the model A is marginal. 
These cases can be considered as estimates for the range of 
  uncertainties of the SN~2013fs mass $10\pm2$\msun\  and the explosion energy   
  $(1.8\pm0.2)\times10^{51}$\,erg.

  
\section{Discussion and Conclusions}
\label{sec:disc}

The primary goal of the paper has been to infer 
  an alternative estimate of the explosion energy in attempt   
  to relax the disparity between the  
  high kinetic energy reported earlier \citep{Yaron_17}
   and the neutrino mechanism. 
The argued origin of the broad \heii\,4686\,\AA\ emission from the 
  fragmented CDS results in the CDS velocity estimate that combined with other 
  observational constraints permits us to infer the energy 
  $\approx 1.8\times10^{51}$ erg for the 10\msun\ ejecta reported 
  in earlier study \citet{Yaron_17} and confirmed here as well. 
Remarkably that the new energy estimate is within the range 
  $\lesssim 2\times10^{51}$\,erg  implied by the neutrino driven explosion  \citep{Janka_17}. 

The large disparity between the present and 
  former energy estimates indicates some systematic effect.
A hint for possible systematic errors related to the energy estimate from 
  the shock breakout analysis is provided by the large difference 
  of the pre-SN radius estimated from the hydrodynamic modelling 
  (1000\rsun) and the analytical consideration (180\rsun) \citep{Yaron_17}. 
Yet, independently of the exact source of systematic error the former energy 
  estimate anyway indicates that the explosion energy of SN~2013fs 
   is likely high compared to other SNe~IIP.
 In that sense the former and the present energy estimate are qualitatively 
   agree with each other.
 It would be interesting to verify whether the 
   phenomenon of the confined dense shell in SNe~IIP is always accompanied by  
   their high explosion energy. 
  
When interpeting the broad \heii\,4686\,\AA\ emission on day 2.42 we omitted 
  the component with the intermediate FWHM of $\approx 1000$\kms.
Although this component is not involved directly in our model for the broad \heii\ line
  we point out sensible explanations for this feature.
The relatively low velocity of the line-emitting gas of the intermediate component 
   suggests that this line is related to the CS gas.  
On the other hand, the absence of a narrow component (FWHM $\sim 100$\kms), unlike, 
  e.g.,that 
  in the H$\beta$ (Fig. \ref{fig:he}), indicates that only a high velocity CS 
  gas is involved in the emission.
Two conjectures are conceivable for the origin of this high velocity CS gas: 
 (i) the preshock  wind accelerated by the supernova radiation; (ii) the accelerated fragments of crushed CS clouds engulfed by the forward shock.
These possibilities should be discriminated via the detailed modelling.
  
At the moment we aware of at least three well observed SNe~II with     
  the estimated extent of the confined dense CS shell ($R_{ds}$): SN 1998S with 
   $R_{ds} \sim 10^{15}$ cm \citep{Chu_01}, SN~2013cu with 
  $R_{ds} \sim (4-7)\times10^{14}$ cm \citep{Groh_14}, and SN~2013fs 
  with $R_{ds} \sim 5\times10^{14}$ cm \citep{Yaron_17}.  
The extension of dense CS shell of another SN~2006bp showing flash spectrum can be estimated 
  based on the fact that the narrow \ha\ in SN~2006bp disappeared betweed days 3 and 5  \citep{Quimby_07}. 
The broad \heii\,4686\,\AA\ in SN~2006bp on day 2 implies the CDS velocity of $\approx 16000$\kms. 
With this average velocity and the lifetime of narrow \ha\ of 4 day the  
  extension of the confined CS shell turns out to be $R_{ds} \approx 5.5\times10^{14}$ cm, comparable with the above three supernovae.
To summarize, in four discussed SNe~II the extent of the confined CS shell  
  falls in the range of $\sim (4-10)\times10^{14}$ cm. 
Among these SNe the wind velocity is well determined only for  SN~1998S  
  in which case $u \approx 30$\kms\ \citep{Chu_02}. 
A reasonable assumption is that  other three SNe have the similar wind speed 
  in which case the strong mass loss occurs between 4 and 10 yrs prior to the core collapse. 
  
The mechanism operating in the pre-collapse core that brings about 
  a vigorous mass loss at this stage has yet to be understood.
Depending on the prerequisites (e.g., mass, rotation, etc.) the mechanism probably 
  operates with different efficiency marked by the fact that only about 
    1/5 of SNe~II demonstrate flash spectra \citep{Khazov_16}.

\section*{Acknowledgements}
I thank V. P. Utrobin for discussions.

\bibliographystyle{mnras}
\bibliography{sn13fs_ref} 
\bsp	
\label{lastpage}
\end{document}